\documentclass[a4paper,conference]{IEEEtran}
\IEEEoverridecommandlockouts
\usepackage{cite}
\usepackage{amsmath,amssymb,amsfonts}
\usepackage{amsthm}
\usepackage{algorithmic}
\usepackage{graphicx}
\usepackage{textcomp}
\usepackage{xcolor}
\usepackage{colortbl}
\usepackage{tabularx}
\usepackage{booktabs}
\usepackage{multirow}
\usepackage{subfig}
\usepackage{balance}
\usepackage{tikz}

\usepackage{mathrsfs}
\usepackage{bm} 
\usepgflibrary{plotmarks} 
\usepackage{pgfplots}
\usepgfplotslibrary{colorbrewer}
\usepgfplotslibrary{external}
\usepgfplotslibrary{groupplots}
\usepgfplotslibrary{ternary}
\usepgfplotslibrary{fillbetween}
\usetikzlibrary{fadings}

\usetikzlibrary{shapes}
\usetikzlibrary{spy}
\usetikzlibrary{arrows,decorations.pathmorphing,backgrounds,positioning,fit,petri, intersections}
\usetikzlibrary{automata}
\usepackage{etoolbox}
\robustify{\mathbb}
\robustify{\mathrm}
\robustify{\mathcal}
\robustify{\{}
\robustify{\}}
\definecolor{Blue}{RGB}{33,26,82}

\tikzset{>=latex,radiation/.style={{decorate,decoration={expanding waves,angle=90,segment length=2pt}}},
         pics/antenna/.style={
        code={\tikzset{scale=0.1}
            \draw[semithick] (0,0) -- (1,4);% left line
            \draw[semithick] (3,0) -- (2,4);% right line
            \draw[semithick] (0,0) arc (180:0:1.5 and -0.5);
            \node[inner sep=4pt] (circ) at (1.5,5.5) {};
            \draw[semithick] (1.5,5.5) circle(8pt);
            \draw[semithick] (1.5,5.5cm-8pt) -- (1.5,4);
            \draw[semithick] (1.5,4) ellipse (0.5 and 0.166);
            \draw[semithick,radiation,decoration={angle=45}] (1.5cm+8pt,5.5) -- +(0:2);
            \draw[semithick,radiation,decoration={angle=45}] (1.5cm-8pt,5.5) -- +(180:2);
  }},
    pics/phone/.style={
  code={
  \node[inner sep=1pt] (circ) at (0,0.42) {};
  \node[minimum size=0.42, circle] (circle) at (0,0.21) {};
    \node[inner sep=1pt] (cent) at (0,0.21) {};
  \filldraw[fill=white, rounded corners=1pt,thick]  (-0.13,0) rectangle (0.13,0.42);
      \draw(-0.13,0.09) --++(0.26,0)++(0,0.26)--++(-0.26,0);
  \filldraw(0,0.04) circle (0.01);
  }},
  pics/rphone/.style={
  code={
  \node[inner sep=0pt] (circ) at (-0.34,0.65) {};
  \filldraw[fill=white, rounded corners=1pt,thick]  (-0.13,0) rectangle (0.13,0.42);
      \draw(-0.13,0.09) --++(0.26,0)++(0,0.26)--++(-0.26,0);
  \filldraw(0,0.04) circle (0.01);
\draw[radiation,decoration={angle=45}] (0.13,0.44) --++(45:0.3);
  }},
  pics/lphone/.style={
  code={
  \node[inner sep=0pt] (circ) at (-0.34,0.65) {};
  \filldraw[fill=white, rounded corners=1pt,thick]  (-0.13,0) rectangle (0.13,0.42);
      \draw(-0.13,0.09) --++(0.26,0)++(0,0.26)--++(-0.26,0);
  \filldraw(0,0.04) circle (0.01);
\draw[radiation,decoration={angle=45}] (-0.13,0.44) --++(135:0.3);
  }},
  pics/eNB/.style={
  code={
  \draw[name path=left, thick](0,0)--(0,0.8) -- (-0.2,0.1) ;
\draw[name path=right, thick](0.2,0.1)--(0,0.8)--(0,0.9);
    \node[minimum size=0.75cm, circle] (circ) at (0,0.9) {};
\foreach \x in{0,0.1,...,0.8}
{
\path [name path=second](-0.2,0.2+\x) --(0,0.1+\x)--(0.2,0.2+\x)--(0,0.3+\x)--cycle;
\draw[name intersections={of=second and right}](0,0.1+\x)--(intersection-1);
\draw[name intersections={of=second and left}](0,0.1+\x)--(intersection-2);
}
\draw[radiation,decoration={angle=45}] (0.05,0.9) --++(0:0.2);
\draw[radiation,decoration={angle=45}] (-0.05,0.9) --++(180:0.2);
  }},
  battery/.pic={
  code={
  \filldraw[fill=white,very thick] (0,0) rectangle (0.5,1.5);
\filldraw[fill=white,very thick] (0.1,1.5) rectangle (0.4,1.55);
\filldraw[fill=white,very thick] (0.2,1.55) rectangle (0.3,1.65);
\filldraw[red!80!black] (0.05,0.05) rectangle (0.45,0.25); 
  }}
}
\tikzset{
/pgfplots/colormap/viridis,
        COLOR/.style={
            index of colormap={#1},
        },
    }

    \tikzstyle{every state}=[circle, very thick,align=center, minimum size=14pt, font =\footnotesize]

\tikzset{master/.style={draw=Blue, fill=Blue, text=white}}

\tikzset{oldn/.style={draw=black, fill=none,font=\scriptsize}}
\tikzset{newn/.style={ draw=Greys-E, fill=Greys-E, font=\scriptsize}}

\pgfplotsset{
compat=newest,  %axis line style={|-|},
major grid style={very thin, Greys-E},
clip=false,
width=0.8\columnwidth, 
height=1.6in, 
scale only axis, 
scaled ticks=false, 
ylabel style={align=center}, 
xlabel style={align=center}, 
every axis legend/.append style={column sep=0.05cm, fill=white, legend cell align=left, font = \footnotesize, align=left},
every tick label/.append style = {font = \footnotesize},
axis background/.append style={fill=white},
every axis plot/.append style={mark options=solid, font = \footnotesize, thick},
every node near coord/.append style={
    /pgf/number format/fixed zerofill,
    /pgf/number format/precision=3, anchor=west
}
}

\pgfplotsset{
  tufte axes/.style =
    {
ymajorgrids=true,
axis line shift=5pt,
ylabel shift=5pt,
xlabel shift=5pt,
}
}

\pgfkeys{/pgf/number format/.cd,fixed relative, precision=5, set thousands separator={}}

\pgfplotscreateplotcyclelist{markers}{%
{mark=x},
{mark=+},
{mark=Mercedes star},
{mark=Mercedes star flipped},
{mark=star}}

 \pgfplotscreateplotcyclelist{lines}{%
	{-},
	{densely dashed},
	{densely dotted},
	{densely dashdotted},
	{densely dashdotdotted},}

\pgfplotscreateplotcyclelist{LinesMarks}{%
	{-},
	{densely dashed, mark=x},
	{densely dotted, mark=+},
	{densely dashdotted, mark=Mercedes star}
	{densely dashdotdotted, mark=Mervedes star inverted}
}

\pgfplotscreateplotcyclelist{ShadesofGrey}{
	{Greys-M},
	{Greys-K},
	{Greys-J},
	{Greys-G},
	{Greys-F}
}

\pgfplotscreateplotcyclelist{ShadesofGreyLines}{
	{Greys-M, -},
	{Greys-L, densely dashed},
	{Greys-K, densely dotted},
	{Greys-J, densely dashdotted},
	{Greys-I, densely dashdotdotted},
	{Greys-H, -}}

\pgfplotscreateplotcyclelist{ShadesofGreyLinesInv}{
	{Greys-H, -},
	{Greys-I, densely dashdotdotted},
	{Greys-J, densely dashdotted},
	{Greys-K, densely dotted},
	{Greys-L, densely dashed},
	{Greys-M, -}}

\pgfplotscreateplotcyclelist{YlGnBu}{
	{YlGnBu-M},
	{YlGnBu-K},
	{YlGnBu-I},
	{YlGnBu-G},
	{YlGnBu-E},
}

\pgfplotscreateplotcyclelist{YlGnBu4}{
	{YlGnBu-M},
	{YlGnBu-J},
	{YlGnBu-H},
	{YlGnBu-F}
}

\pgfplotscreateplotcyclelist{YlGnBuInv}{
	{YlGnBu-G},
	{YlGnBu-I},
	{YlGnBu-K},
	{YlGnBu-M}
}

\pgfplotscreateplotcyclelist{vir6}{
	{COLOR=0},
	{COLOR=3},
	{COLOR=6},
	{COLOR=9},
	{COLOR=12},
	{COLOR=15}
}

\pgfplotscreateplotcyclelist{vir6inv}{
	{COLOR=15},
	{COLOR=12},
	{COLOR=9},
	{COLOR=6},
	{COLOR=3},
	{COLOR=0}
}

\pgfplotscreateplotcyclelist{vir5}{
	{COLOR=0},
	{COLOR=4},
	{COLOR=7},
	{COLOR=10},
	{COLOR=13}
}

\pgfplotscreateplotcyclelist{vir4}{
	{COLOR=0},
	{COLOR=5},
	{COLOR=9},
	{COLOR=14}
}

\pgfplotscreateplotcyclelist{Q3}{
	{black},
	{Greens-G},
	{Blues-K}
}

\pgfplotscreateplotcyclelist{Dark3}{
	{Dark2-A},
	{Dark2-B},
	{Dark2-C}
}

\definecolor{Dark}{RGB}{37,52,148}

\newcommand{\dd}{\mathop{}\!\mathrm{d}}
\newcommand{\astretch}[1]{\renewcommand*{\arraystretch}{#1}}
\newcommand{\wh}[1]{$\widehat{\text{#1}}$}
\theoremstyle{definition}

\DeclareMathOperator{\E}{\mathbb{E}}
\def\BibTeX{{\rm B\kern-.05em{\sc i\kern-.025em b}\kern-.08em
    T\kern-.1667em\lower.7ex\hbox{E}\kern-.125emX}}

\begin{document}
\title{Wireless Mesh Networking with Devices Equipped with Multi-Connectivity
\thanks{The research of I. Leyva-Mayorga and Petar Popovski was supported by Intel Deutschland through the Multi-RAT mesh project.}
}

\author{
\IEEEauthorblockN{
Israel Leyva-Mayorga\IEEEauthorrefmark{1}, Radoslaw~Kotaba\IEEEauthorrefmark{1}, Fresia~Maria\IEEEauthorrefmark{2}, and Petar~Popovski\IEEEauthorrefmark{1}
}
\IEEEauthorblockA{\IEEEauthorrefmark{1}Connectivity Section, Department of Electronic Systems, Aalborg University, Denmark\\
email:\{ilm, rak, petarp\}@es.aau.dk\\%
}
\IEEEauthorblockA{\IEEEauthorrefmark{2}Intel Deutschland \\
Munich, Germany\\
email: maria.fresia@intel.com}}

\maketitle

\begin{abstract}
Wireless connectivity is rapidly becoming ubiquitous and affordable. As a consequence, most wireless devices are nowadays equipped with multi-connectivity, that is, availability of multiple radio access technologies (RATs). Each of these RATs has different characteristics that can be suitably utilized for different connectivity tasks. For example, a long-range low-rate RAT can be used for topology management and coordination, whereas a short-range high-rate RAT for data transmission. In this paper, we introduce a distributed consensus protocol for the hierarchical organization of Wireless Mesh Networks (WMNs) with devices using multiple RATs. Our protocol considers three hierarchical roles after the initial setup: Master, cluster head (CH), and cluster member (CM). The Master coordinates the use of all RATs, whereas the CHs coordinate all but the RAT with the longest transmission range. The initial setup takes place immediately after powering on the devices, after which the devices self-organize in a distributed manner by means of a consensus to elect the Masters and CHs. The resulting interconnected structure is 
based on the connectivity graphs created with the different RATs. The distributed consensus protocol operates with a minimal amount of network information and demonstrates high networking performance.
\end{abstract}


\section{Introduction}
%\ilmcom{Distributed consensus protocol $\rightarrow$ DCP?}
%\ilmcom{Is it hierarchical structure or architecture?}
As technology evolves, incorporating multiple radio access technologies (RATs) in a single device has gone from being a mere commodity with minimal added benefits to being a predominant, inexpensive, and valuable feature. This is usually referred to as \emph{multi-connectivity} and has been seen as a key enabler of ultra-reliable wireless communication \cite{nielsen2017ultra,Wolf2019}. 
For instance, most of the current handheld devices incorporate a mix of 4G/5G (cellular), WiFi, Bluetooth, and low-power wide-area (LPWA) technologies. Each of these RATs presents a unique blend of characteristics such as power consumption, maximum coupling loss (MCL), carrier frequency, and bandwidth, among others. The combination of these characteristics is, in the end, reflected in the transmission range, defined as the maximum distance at which communication is reliable, and in the achievable data rate. The diversity provided by multiple RATs with different transmission ranges and data rates is particularly appealing to wireless mesh networks (WMNs), where numerous devices organize in a random topology to communicate with each other.

Single-RAT WMNs were a hot research topic in the early 2000s and significant advances were made regarding their performance limits. For instance, experimental results demonstrated that there exists a horizon in the number of hops and number of devices for functional wireless networking~\cite{Tschudin2005}. As the number of devices increases, routing becomes more complex, medium access becomes inefficient, the overhead of coordination grows, and resource allocation has to be relaxed. 

The performance of a WMN is limited by coverage if the RATs that are used have relatively short transmission ranges, thereby necessitating the use of multi-hop routes. Conversely, the performance of a WMN that uses RATs with relatively long transmission ranges (e.g. unlicensed spectrum in sub-$1$~GHz bands) is limited by interference. That is, any given destination is within one or a few hops from any given source. 
Gupta and Kumar~\cite{Gupta2000} concluded that the optimal throughput of an interference-limited WMN with $N$ devices scales as $\mathcal{O}\left(\sqrt{N}\right)$. Later, \"{O}zg\"{u}r et al.~\cite{Ozgur2007} showed that the optimal throughput in such a WMN scales linearly with $N$ if and only if hierarchical cooperation is used to support multiple simultaneous long-range transmissions.

The use of devices equipped with multi-connectivity brings a fresh breeze to the concept of WMN and it can greatly exceed the performance limits imposed by each individual RAT, provided that the RATs are efficiently coordinated. For instance, the flexible use of long- and short-range RATs helps maintain low levels of interference in the long-range RAT while allowing for spatial reuse in the short-range. In addition, having multiple RATs further increases the micro-diversity of the WMN, that is, the spatial and frequency diversity that can be greatly effective to withstand small-scale fading~\cite{Wolf2019}. 

On the downside, multi-connectivity increases the number of degrees of freedom to the already-complex challenges in WMNs. For example, Borst et al.~\cite{Borst2017} observed that efficient path selection algorithms for multi-RAT networks must be: 1) dynamic, 2) able to predict the channel conditions, and 3) designed to allow for fair resource sharing. These conclusions were drawn from a small single-hop multi-RAT setup. In larger multi-RAT WMNs, machine learning techniques may be a better option for routing and coordination, where the complexity of these challenges increases rapidly. Nevertheless, these approaches have been only initially investigated for single-RAT WMNs~\cite{Karunaratne2019}.

In this paper, we present a distributed consensus protocol to support dynamic networking in multi-RAT WMNs. To the best of our knowledge, ours is the only distributed protocol in the literature that exploits the different characteristics of individual RATs to induce a hierarchical network architecture.
Three different hierarchy roles are assigned during the initial setup of the WMN (from highest to lowest): Master, cluster head (CH), and cluster member (CM). The Master coordinates all the RATs within its cluster and the use of the RAT with the longest transmission range in the whole network; hereafter, we refer to this as the long-range RAT and to the rest as the short-range RATs. The Master is also in charge of allocating resources to the clusters, selecting the appropriate medium access protocol, and scheduling the subsequent phases of the network. 
The CHs coordinate the use of all the RATs and allocate resources within their cluster. This includes coordinating the use of the long-range RAT based on the instructions received from the M, in addition to coordinating the use of the short-range RATs. In other words, the Master is a CH that also performs other essential management tasks in the network. CMs have the lowest hierarchy after setup and their communication is subject to the coordination of the Master and their CH. Each node is associated with a CH and with a Master.

\begin{figure}[t]
\centering
\subfloat[]{\includegraphics{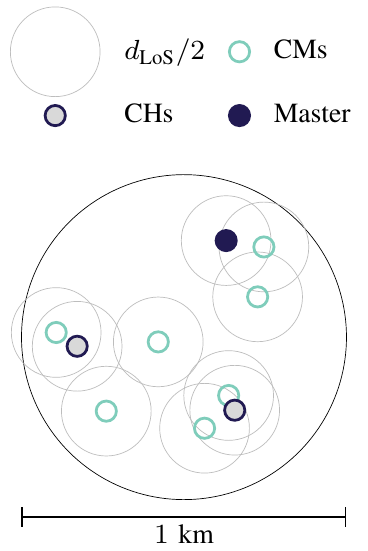}\label{fig:example_a}}\hfil
\subfloat[]{
\begin{tikzpicture}[auto, every node/.append style={inner sep=2pt, minimum size=2.25pt}]
\coordinate (S) at (0,0);
\def\g{36}
\def\x{0.8}
\coordinate (A) at ($(S) +(1*\g:\x*1cm)$);
\coordinate (B) at ($(S) +(2*\g:\x*1cm)$);
\coordinate (C) at ($(S) +(7*\g:\x*1cm)$);
\coordinate (D) at ($(S) +(0*\g:\x*1cm)$);
\coordinate (E) at ($(S) +(8*\g:\x*1cm)$);
\coordinate (F) at ($(S) +(4*\g:\x*1cm)$);
\coordinate (G) at ($(S) +(5*\g:\x*1cm)$);
\coordinate (H) at ($(S) +(3*\g:\x*1cm)$);
\coordinate (I) at ($(S) +(6*\g:\x*1cm)$);
\coordinate (J) at ($(S) +(9*\g:\x*1cm)$);
\path[thick, Greys-G]
(A) edge (B)
(A) edge (D)
(A) edge (H)

(B) edge (D)
(B) edge (G)

(C) edge (E)
(C) edge (J)

(D) edge (H)

(E) edge (J)

(F) edge (G)
(F) edge (I)

(G) edge (I);

\filldraw[thick, fill=white, draw=YlGnBu-F] (A) circle (2.25pt);
\filldraw[thick, fill=white, draw=YlGnBu-F] (B) circle (2.25pt);
\filldraw[thick, fill=white, draw=YlGnBu-F] (C) circle (2.25pt);
\filldraw[thick, draw=Blue, fill=Greys-D] (D) circle (2.25pt);
\filldraw[thick, fill=Blue, draw=Blue] (E) circle (2.25pt);
\filldraw[thick, fill=white, draw=YlGnBu-F] (F) circle (2.25pt);
\filldraw[thick, fill=white, draw=YlGnBu-F] (G) circle (2.25pt);
\filldraw[thick, fill=white, draw=YlGnBu-F] (H) circle (2.25pt);
\filldraw[thick, draw=Blue, fill=Greys-D] (I) circle (2.25pt);
\filldraw[thick, fill=white, draw=YlGnBu-F] (J) circle (2.25pt);

\def\d{0}
\def\y{-2.2}

\coordinate (A) at ($(S) +(1*\g:\x*1cm)+(\d,\y)$);
\coordinate (B) at ($(S) +(2*\g:\x*1cm)+(\d,\y)$);
\coordinate (C) at ($(S) +(7*\g:\x*1cm)+(\d,\y)$);
\coordinate (D) at ($(S) +(0*\g:\x*1cm)+(\d,\y)$);
\coordinate (E) at ($(S) +(8*\g:\x*1cm)+(\d,\y)$);
\coordinate (F) at ($(S) +(4*\g:\x*1cm)+(\d,\y)$);
\coordinate (G) at ($(S) +(5*\g:\x*1cm)+(\d,\y)$);
\coordinate (H) at ($(S) +(3*\g:\x*1cm)+(\d,\y)$);
\coordinate (I) at ($(S) +(6*\g:\x*1cm)+(\d,\y)$);
\coordinate (J) at ($(S) +(9*\g:\x*1cm)+(\d,\y)$);
\path[thick, Blue]
(A) edge (B)
(A) edge (C)
(A) edge (D)
(A) edge (E)
(A) edge (F)
(A) edge (G)
(A) edge (H)
(A) edge (I)
(A) edge (J)

(B) edge (C)
(B) edge (D)
(B) edge (E)
(B) edge (F)
(B) edge (G)
(B) edge (H)
(B) edge (I)
(B) edge (J)

(C) edge (D)
(C) edge (E)
(C) edge (F)
(C) edge (G)
(C) edge (H)
(C) edge (I)
(C) edge (J)

(D) edge (E)
(D) edge (F)
(D) edge (G)
(D) edge (H)
(D) edge (I)
(D) edge (J)

(E) edge (F)
(E) edge (G)
(E) edge (H)
(E) edge (I)
(E) edge (J)

(F) edge (G)
(F) edge (H)
(F) edge (I)
(F) edge (J)

(G) edge (H)
(G) edge (I)
(G) edge (J)

(H) edge (I)
(H) edge (J)

(I) edge (J)
;

\filldraw[thick, fill=white, draw=YlGnBu-F] (A) circle (2.25pt);
\filldraw[thick, fill=white, draw=YlGnBu-F] (B) circle (2.25pt);
\filldraw[thick, fill=white, draw=YlGnBu-F] (C) circle (2.25pt);
\filldraw[thick, draw=Blue, fill=Greys-D] (D) circle (2.25pt);
\filldraw[thick, fill=Blue, draw=Blue] (E) circle (2.25pt);
\filldraw[thick, fill=white, draw=YlGnBu-F] (F) circle (2.25pt);
\filldraw[thick, fill=white, draw=YlGnBu-F] (G) circle (2.25pt);
\filldraw[thick, fill=white, draw=YlGnBu-F] (H) circle (2.25pt);
\filldraw[thick, draw=Blue, fill=Greys-D] (I) circle (2.25pt);
\filldraw[thick, fill=white, draw=YlGnBu-F] (J) circle (2.25pt);

\draw[thick,Greys-G] (-1.3, 1.5*1cm+1.5em) --++(0.4,0)node[right, inner sep=3pt, black]{short-range link};
\draw[thick,Blue] (-1.3, 1.5) --++(0.4,0)node[right, inner sep=3pt, black]{long-range link};

\end{tikzpicture}\label{fig:example_b}}
\caption{Hierarchical architecture of a WMN with two RATs after initial setup with our distributed consensus protocol. (a) Spatial distribution and (b) connectivity graphs with a short-range (above) and a long-range RAT (below).}
\label{fig:example}
\end{figure}
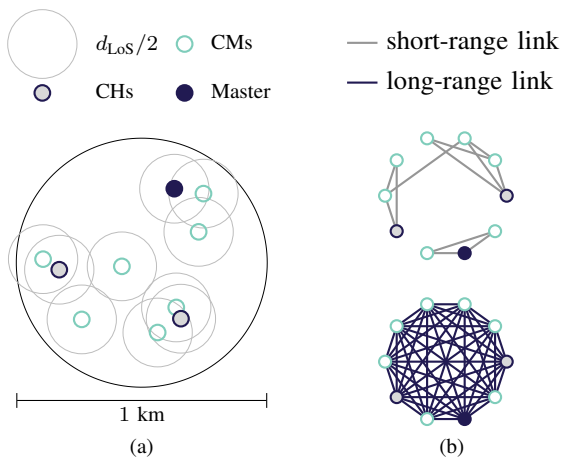

The benefits of having a hierarchical architecture, where some of the nodes coordinate the different RATs, are greatly appealing for multi-RAT WMNs. However, these have not yet been fully exploited. For instance, the well known Optimized Link State Routing (OLSR)~\cite{Clausen2003} supports multiple RATs and defines two levels of hierarchy: normal nodes and multi-point relays (MPRs), which are selected for each RAT. Only MPRs are allowed to forward the data generated by neighboring devices. This approach is efficient in coverage-limited WMNs, but not in interference-limited WMNs, where devices form complete connectivity graphs.

Fig.~\ref{fig:example} shows the roles assigned by our distributed consensus protocol after the initial setup in a multi-RAT WMN. Two RATs were considered and the discs depict half of the line of sight (LoS) distance $d_\text{LoS}/2$. The latter is calculated from the channel model described in Section~\ref{sec:model}. Thus, two devices can communicate with the short-range RAT w.h.p. whenever two discs are overlapped. The transmission range for the long-range RAT is sufficient to communicate the Master with all the CHs.

As the connectivity graphs for the two RATs in Fig.~\ref{fig:example_b} illustrate, the communication between two devices exclusively with the short-range RAT may be complicated and inefficient due to the number of hops required, or even impossible. On the other hand, the levels of interference may become excessive if the long-range RAT is not effectively coordinated. This latter problem persists if OLSR were implemented in the devices of Fig.~\ref{fig:example} because the devices form a complete network graph, where there are no two-hop neighbors. Hence, no device is selected as MPR. 

The main design goal of our distributed consensus protocol is to assign the roles defined above to ensure that the use of every RAT $r$ in every device in the network is coordinated by at least one device. Extensive simulations and analysis show that our distributed consensus protocol achieves the desired hierarchical architecture in relatively dense WMNs with minimal network information and communication overhead. In particular, out protocol greatly reduces the utilization of the long-range RAT when compared to neighbor discovery with the single-RAT only. This is of utmost importance because unlicensed frequency bands for long-range transmission are usually subject to duty cycle.  %We have evaluated the overhead of our protocol in terms of the utilization of each RAT that is needed to reach the distributed consensus. These include the handshakes needed to create the network topology, to propose and accept role changes, and to inform the devices about the accepted changes.

The rest of the paper is organized as follows. Section~\ref{sec:model} presents the system model. Next, Section~\ref{sec:analysis} showcases the benefits of effectively coordinated multi-RAT WMNs. Section~\ref{sec:protocol} presents our distributed consensus protocol for multi-RAT WMNs. Then, Section~\ref{sec:results} presents results on the stability of the resulting hierarchical architecture and on the efficiency of our protocol. Section~\ref{sec:conc} presents the relevant conclusions and future work.

%%%%%%%%%%%%%%%%%%%%%%%%%%%%%%%%%%%%%%%%%%%%%%%%%%%%%%%%%%%%%%%%%%%%%%%%%%%%%%%%%%%%%%%%%%%%%%%%%%%%%%%%%%%%%%%%%%%%%%%%%%%%%%%%%%%%%%%%%%%%%%%%
\section{System model}
\label{sec:model}

%Let $X$ be the random variable that represents the time at which an event is triggered. Building on this, $X_\text{on}^{(i)}$, $X_\text{CH}^{(i)}$, and $X_\text{M}^{(i)}$ be the RVs that represent the time at which device $i$ powers on, become CH, and become the Master, respectively. 

We consider a set of devices that are distributed within an area $A$. These devices communicate through two different RATs $\mathcal{R}=\{1,2\}$. The short-range RAT $r=1$ operates in the $2.4$~GHz band and the long-range RAT $r=2$ operates in the $868$~MHz ISM band. Note that the use of this unlicensed band is restricted by a duty cycle that varies depending on the selected sub-band. We assume that the communication with RAT $r=2$ takes $\rho$ times more time than with RAT $r=1$. This includes all the aspects of communication, such as the medium access and data transmission.

A large-scale fading channel is considered. The total loss in decibels between two devices communicating through RAT $r$ with carrier frequency $f_r$ and whose euclidean distance is $d$ is a random variable (RV) denoted by $L(d,f_r)$. It is calculated according to a site-general path loss model for terminals located near street level~\cite[Section 4.3.1]{ITU2017}. Specifically, the median path loss under line-of-sight (LoS) is
% A large-scale fading channel is considered. We denote $L(d,r)$ as the random variable (RV) of the loss in decibels between two devices that communicate through RAT $r$ and whose euclidean distance is $d$. The distribution of $L(d,r)$ is calculated by a site-general path loss model for terminals located near street level~\cite[Section 4.3.1]{ITU2017}. Specifically, the median path loss under line-of-sight (LoS) is
\begin{equation}
    L_\text{LoS}(d,f)=32.45 +20\log_{10}(f)+ 20\log_{10}\left(d\right) + \Delta L_\text{LoS}
\end{equation}
and the LoS location correction, which depends on the type of urban environment $L_\text{urban}$, is
\begin{multline}
L_\text{NLoS}(d,f)=9.5 +45\log_{10}(f) + 40\log_{10}\left(d\right)\\+L_\text{urban} + \Delta L_\text{NLoS}.
\end{multline}
The location correction percentage $p$ determines the LoS distance as
\begin{equation}
d_\text{LoS}= 212(\log_{10} p)^2 - 64 \log_{10}{p}
\end{equation}
From there, the average loss at a given distance is
\begin{equation}
    \overline{L}(d,f)=\begin{cases}
    L_\text{LoS}(d,f) & \text{if } d<d_\text{LoS}\\
    L_\text{NLoS}(d,f) & \text{if } d>d_\text{LoS}+w\\
    g(d,w) &  \text{otherwise}
    \end{cases}
\end{equation}
where $w$ is the width of the transition zone between LoS and NLoS given in meters, and $g(d,w)$
%\begin{IEEEeqnarray}{rCl}
 %    g(d,w) &=& L_\text{NLoS}(d_\text{LoS},f)+\bigg(\frac{d- d_\text{LoS}}{w}\bigg)\IEEEnonumber\\
  %   &&\times\big(L_\text{NLoS}(d_\text{LoS}+w,f)-L_\text{NLoS}(d_\text{LoS},f)\big)
%\end{IEEEeqnarray}
is the linear interpolation within $d_\text{LoS}<d<d_\text{LoS} + w$.

The total loss in decibels is
\begin{equation}
    L(d,f_r)= \overline{L}(d,f_r)+\chi~\text{dB}
\end{equation}
where $\chi$ is the Gaussian RV with zero mean and standard deviation $\sigma$ denoting the shadow fading. Hence, the outage probability at a distance $d$ is given as
\begin{IEEEeqnarray}{rCl}
    p_\text{out}(d,r)&=&\Pr\left[L\left(d,f_r\right)>L_{\max}(r)\right]\IEEEnonumber\\
    &=& Q\left(\frac{\overline{L}(d,f_r)-L_{\max}(r)}{\sigma}\right)
\end{IEEEeqnarray}
where $Q(z)=\frac{1}{\sqrt{2 \pi}}\int_z^\infty e^{-u^2/2}\dd u$. 

\begin{figure}
    \centering
    \includegraphics{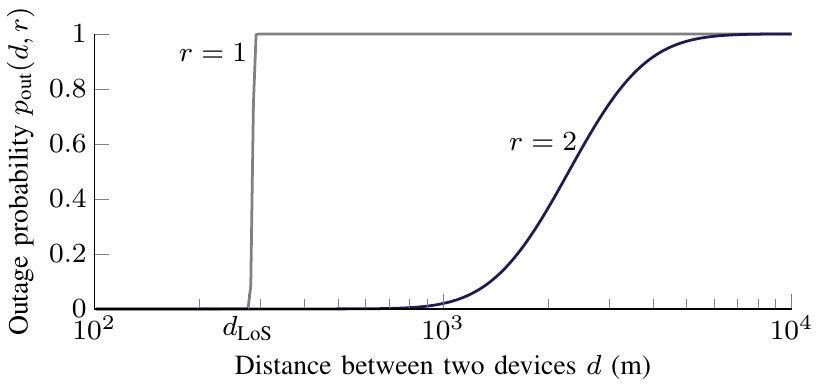}
    \caption{Outage probability $p_\text{out}(d,r)$ for $\mathcal{R}=\{1,2\}$.}
    \label{fig:channel}
\end{figure}

We assume a typical value for the width of the transition zone $w=20$~m and for the standard deviation of the log-normal shadowing $\sigma=7$~dB, along with $p=0.1$ and $L_\text{urban}=6.8$~dB, which give $d_\text{LoS}=276$~m, $\Delta L_\text{LoS}=-7.9$~dB, and $\Delta L_\text{NLoS}=-9$~dB. Furthermore, we select typical values for $L_{\max}(1)= 105$ and $L_{\max}(2)=154$. Fig.~\ref{fig:channel} illustrates the outage probability for each of the considered RATs given these parameter settings.

\iffalse
The rest of the relevant parameter settings are listed in Table~\ref{tab:param}.
\begin{table}[t]
    \centering
    \astretch{1.3}
    \caption{Parameter settings for the RATs.}
    \begin{tabularx}{\columnwidth}{@{}Xlll@{}}
        \toprule
        Parameter &Symbol & \multicolumn{2}{c}{Setting} \\\cmidrule{3-4}
        && $r=1$ & $r=2$\\
        \midrule
       LoS distance [m] & $d_\text{LoS}$ &\multicolumn{2}{c}{$ 276$}\\
       Standard deviation of the shadow fading~[dB] & $\sigma$& \multicolumn{2}{c}{$7$}\\
       Carrier frequency [GHz]& $f_r$ &$2.4$ 
       &$0.868$\\
        Maximum coupling loss (MCL) [dB] & \\
      \bottomrule
    \end{tabularx}
    \label{tab:param}
\end{table}
\fi
Throughout this paper, we assume the coherence time of the channel is sufficiently long so that communication between $i$ and $j$ is possible if the channel is not in outage. This occurs with probability
\begin{equation}
    p_{ij}^{(r)}=p(d,r)=1-p_\text{out}(d,r)
\end{equation}
The devices use a simple neighbor discovery algorithm that is based on a random access protocol. In particular, a connection between two devices $i$ and $j$ is established with RAT $r$ if the exchange is successfully completed. Hence, the probability that a connection between $i$ and $j$ is established is $p_{ij}^{(r)}$.  

Building on the characteristics described above, the connectivity in the short-range RAT $r=1$ could be approximated as a Gilbert random disk graph with disc radius $\approx d_\text{LoS}/2$ as shown in Fig.~\ref{fig:example}.

In the following, we present a simple example to illustrate the benefits of multi-RAT WMNs w.r.t. single-RAT WMNs.

%in which a hierarchical structure greatly reduces the complexity of resource allocation and coordination.
%%%%%%%%%%%%%%%%%%%%%%%%%%
\section{Benefits of multi-RAT WMNs}
\label{sec:analysis}
\begin{figure}[t]
    \flushright
    \tikzset{node distance=1.1cm}
    \subfloat[]{
\begin{tikzpicture}[auto, every node/.append style={inner sep=2pt}]
\coordinate (S) at (0,0);
\node[state] (A) at (S) {$1$};
\node[state] (B) [right of=S]{$2$};
\node[state] (C) [right of=B]{$3$};
\node[state] (D) [right of=C]{$4$};
\node [above of=S, node distance=0.5cm]{source};
\node [above of=D, node distance=0.5cm]{destination};

\path[thick, Greys-G,->]
(A) edge (B)
(B) edge (C)
(C) edge (D)
;
\coordinate (L) at (4.3,0.25);
\draw[Greys-G, thick] (L) --++(0.8,0)node[right, inner sep=4pt, black]{$r=1$};

\draw[black, thick,->] ($(L) + (0,-0.5)$) --++(0.8,0)node[right, inner sep=4pt, align=center]{directed edge};

\phantom{\draw[black, thick,<->] ($(L) + (0,-0.5)$) --++(0.8,0)node[right, inner sep=4pt, align=center]{undirected edge};}

\def\y{-0.5}
\draw[|-|]($(A)+(0,\y)$)  -- ($(B)+(0,\y)$)node[midway, anchor=center, fill=white]{\footnotesize $d_{\min}$};
\draw[|-|]($(B)+(0,\y)$)  -- ($(C)+(0,\y)$)node[midway, anchor=center, fill=white]{\footnotesize$d_{\min}$};
\draw[|-|]($(C)+(0,\y)$)  -- ($(D)+(0,\y)$)node[midway, anchor=center, fill=white]{\footnotesize$d_{\min}$};

\end{tikzpicture}\label{fig:analysis_a}}\\
\subfloat[]{
\begin{tikzpicture}[auto, every node/.append style={inner sep=2pt}]
\coordinate (S) at (0,0);
\node[state] (A) at (S) {$1$};
\node[state] (B) [right of=S]{$2$};
\node[state] (C) [right of=B]{$3$};
\node[state] (D) [right of=C]{$4$};
\node [above of=S, node distance=0.8cm]{source};
\node [above of=D, node distance=0.8cm]{destination};

\coordinate (L) at (4.3,0.4);
\draw[Blue, thick] ($(L) + (0,0)$) --++(0.8,0)node[right, inner sep=4pt, black]{$r=2$};

\draw[black, thick,<->] ($(L) + (0,-0.5)$) --++(0.8,0)node[right, inner sep=4pt, align=center]{undirected edge};

\path[thick, bend left, Blue,->]
(A) edge (B)
(A) edge (C.110)
(A) edge (D.110)
(C) edge (D)
;

\path[thick, bend right, Blue,->]
(B) edge[<->, bend left] (C)
(B) edge (D)
;
\end{tikzpicture}\label{fig:analysis_b}}\hfill
\subfloat[]{
\begin{tikzpicture}[auto, every node/.append style={inner sep=2pt}]
\coordinate (S) at (0,0);
\node[state] (A) at (S) {$1$};
\node[state] (B) [right of=S]{$2$};
\node[state] (C) [right of=B]{$3$};
\node[state] (D) [right of=C]{$4$};
\node [above of=S, node distance=0.8cm]{source};
\node [above of=D, node distance=0.8cm]{destination};

\coordinate (L) at (4.3,0.4);
\phantom{\draw[Blue, thick] ($(L) + (0,0)$) --++(0.8,0)node[right, inner sep=4pt, black]{$r=2$};}

\phantom{\draw[black, thick,<->] ($(L) + (0,-0.5)$) --++(0.8,0)node[right, inner sep=4pt, align=center]{undirected edge};}

\path[thick, bend left, Blue,->]
(A) edge (B)
(A) edge (C.110)
(A) edge (D.110)
(C) edge (D)
;

\path[thick, bend right, Blue,->]
(B) edge[<->, bend left] (C)
(B) edge (D)
;

\path[thick, Greys-G,->]
(A) edge (B)
(B) edge[<->] (C)
(C) edge (D)
;
\end{tikzpicture}\label{fig:analysis_c}}
\caption{Data transmission in a linear network with four nodes with (a) a single short-range RAT, (b) a single long-range RAT, and (c) combined short- and long-range RATs (multi-RAT).}
    \label{fig:analysis}
\end{figure}
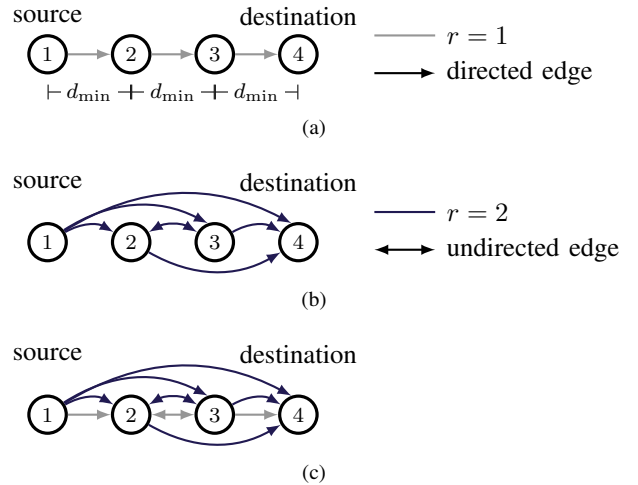

In this section, we provide an example to illustrate the benefits of an effectively coordinated WMN with two RATs with respect to having a single short-range RAT $r=1$ or a single long-range RAT $r=2$. For this, we consider the random graphs depicted in Fig.~\ref{fig:analysis}, where each node (i.e., vertex) represents a communication device. These connectivity graphs represent a linear WMN $\left\{G_{\{r\}}=(V,E_{\{r\}})\right\}$ with set of RATs $\{r\}$, set of vertices $V=\{1,2,3,4\}$, and set of edges $E_{\{r\}}$. Note the undirected (i.e., bidirectional) edges between $2$ and $3$ in Fig.~\ref{fig:analysis_b} and Fig.~\ref{fig:analysis_c}. The inter-node distance is $d_{ij}=\alpha d_{\min}$ for all $|i-j|=\alpha$ and $2d_{\min}>d_\text{LoS}$.
Therefore, nodes can only reach their closest neighbor with $r=1$, but may forward the packets to any node with $r=2$. That is, $p_{i j}^{(1)}=0$ for all $|i-j|>1$ and $p_{i j}^{(2)}>0$ for all $\{i,j\}$. We assume the number of sub-bands in each RAT is sufficient and that the nodes are coordinated to allow multiple simultaneous transmissions.

In the toy example shown here, we analyse the case where node $1$ attempts to communicate with $4$ for a given $\rho$. Hence, communication with $r=1$ takes $1$ time unit and with $r=2$ takes $\rho$ time units. As a result, the graphs illustrated in Fig.~\ref{fig:analysis} are weighted random graphs, where the weight of each edge $e$ is the time needed to communicate through the edge. The performance of the WMNs depicted in Fig.~\ref{fig:analysis} is evaluated in terms of reliability, defined as the probability that node $1$ is able to communicate with $4$, and of the experienced latency.

Let $\mathcal{S}$ be the event that there exists at least one path from node $1$ to $4$. Thus, we denote the reliability with a given set of RATs $\{r\}$ as $\Pr\left[\mathcal{S}\mid \{r\}\right]$. We also denote $P^a$ to be a particular possible path from $1$ to $4$ with set of edges $e(P^a)$. Path $P^a$ exists with probability
\begin{equation}
\Pr\left[P^a\right] = \prod_{ij\in e(P^a)} p_{ij}^{(r)}.
\end{equation}
From there, we calculate 
\begin{equation}
    \Pr\left[\mathcal{S}\mid \{r\}\right]=\Pr\left[\bigcup_{\forall a} P^a\mid \{r\}\right]
    \end{equation}
by taking into account that the possible paths are not mutually exclusive.

Next, let $S$ be the RV of the latency of the communication from $1$ to $4$. We denote the pmf and the CDF of $S$ for a given set of RATs $\{r\}$ as $p_{S}^{(\{r\})}(s)=\Pr\left[S=s\right]$ and $F_{S}^{(\{r\})}(s)=\sum_{k=1}^s p_{S}^{(\{r\})}(k)$, respectively. Recall that the weight of the edges represents the time units needed for communication through an edge, $1$ for $r=1$ and $\rho$ for $r=2$. Thus, $S$ can be calculated as the weight of the shortest weighted path from node $1$ to $4$.   

We start the analysis with the coverage-limited WMN depicted in Fig.~\ref{fig:analysis_a}. Here, only $r=1$ is available, so the data packet must be routed as shown in Fig.~\ref{fig:analysis_a}, which gives
\begin{equation}
    \Pr\left[\mathcal{S}\mid\{1\}\right]=F_S^{(1)}(3)=p_{S}^{(1)}(3)=
    p_{12}^{(1)}\,p_{23}^{(1)}\,p_{34}^{(1)}.
\end{equation}

Next, Fig.~\ref{fig:analysis_b} depicts an interference-limited WMN, where only $r=2$ is available. In this case, direct communication from $1$ to $4$ occurs with probability $p^{(2)}_{14}$. Following a traditional shortest path routing approach no relay devices will be selected, and the reliability of the communication is
$p_S^{(2)}(\rho)=p^{(2)}_{14}$.
Furthermore, the transmissions from $1$ will be treated as interference to nodes $2$ and $3$, which may not be able to communicate until the channel becomes idle again.

Conversely, if nodes $2$ and $3$ are correctly coordinated, these can serve as relays to improve the reliability of the transmission. By doing so, we have
\begin{equation}
     p_S^{(2)}(2\rho)=\left(p^{(2)}_{12}\,p^{(2)}_{24}\left(1-p^{(2)}_{13}\,p^{(2)}_{34}\right) + p^{(2)}_{13}\,p^{(2)}_{34}  \right)\left(1-p^{(2)}_{14}\right) 
\end{equation}
and
\begin{IEEEeqnarray}{rCl}
     p_S^{(2)}(3\rho)&=&p^{(2)}_{23}\left(1-p^{(2)}_{14}\right) \left[p^{(2)}_{12}\,p^{(2)}_{34}\left(1-p^{(2)}_{24}\right)\left(1-p^{(2)}_{13}\right)\right. \IEEEnonumber\\
     &&+\left. p^{(2)}_{13} \,p^{(2)}_{24}\left(1-p^{(2)}_{12}\right) \left(1-p^{(2)}_{34}\right) \right].
\end{IEEEeqnarray}

Naturally, the number of possible paths from $1$ to $4$ for the multi-RAT WMN depicted in Fig.~\ref{fig:analysis_c} is greater than for the cases where one RAT is available. While the increased number of paths may complicate routing and medium access, the hierarchical architecture induced by our protocol can greatly simplify coordination. For example, some links (i.e., edges in the graph $G_\mathcal{R}$) that do not provide a significant benefit may be removed. 

The number of possible paths in combination with the possible coordination approaches requires a detailed analysis that is out of the scope of this paper. Instead, an illustrative lower bound in the performance of $G_\mathcal{R}$ can be obtained by assuming the RATs are used independently as
\begin{equation}
    F_S^{(\mathcal{R})}(s) \geq 1-\left(1-F_S^{(1)}(s)\right)\left(1-F_S^{(2)}(s)\right).
    \label{eq:lowerbound}
\end{equation}

\begin{figure}[t]
\flushright
\subfloat[]{\includegraphics{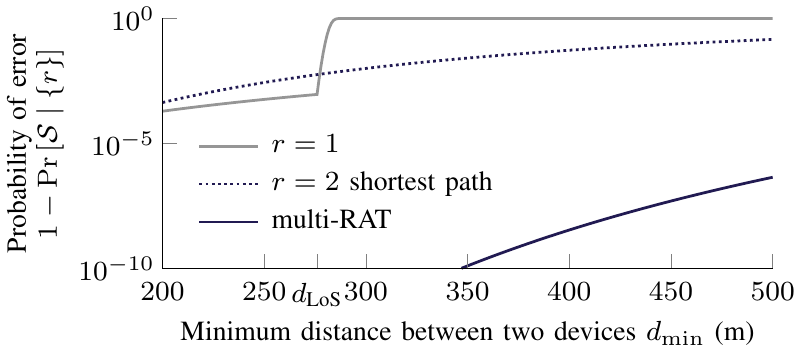}\hspace{2em}\label{fig:reliability}}\\
\subfloat[]{\includegraphics{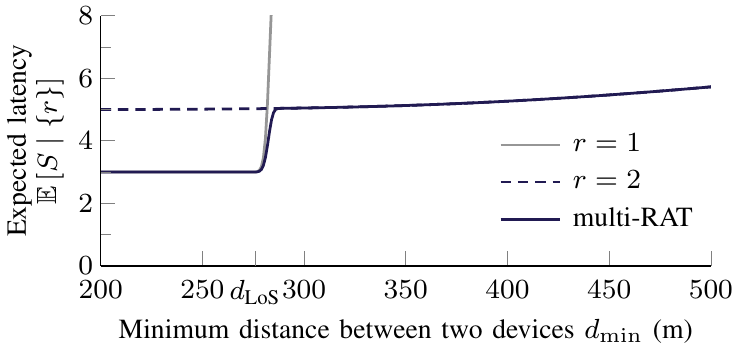}\hspace{2em}\label{fig:latency}}
\caption{(a) Probability of error in logarithmic scale and (b) expected latency for the WMNs depicted in Fig.~\ref{fig:analysis}.}
\label{fig:performance}
\end{figure}

Fig.~\ref{fig:performance} shows the probability of error $1-\Pr\left[\mathcal{S}\mid\{r\}\right]$ in logarithmic scale and the expected latency as a function of the minimum distance between devices. The results illustrate the potential benefits of an effectively coordinated multi-RAT WMN compared to WMNs with a single, short- or long-range, RAT. Note that the lower bound from~\eqref{eq:lowerbound} and $\rho=5$ are used for this analysis. 

As Fig.~\ref{fig:reliability} shows, the reliability of the multi-RAT WMN exceeds that of a WMN with $r=1$ only and that of a WMN with $r=2$ and no relays for large values of $d_{\min}$. Conversely, the reliability of the multi-RAT WMN is similar to that of the WMN with $r=2$, but only if this RAT is effectively coordinated and all possible paths are used. Hence, the latter plot was omitted from Fig.~\ref{fig:reliability}.

Nevertheless, the use of two RATs results in the minimum communication latency, as shown in Fig.~\ref{fig:latency} for all values of $d_{\min}$. In particular, the latency for a WMN with $r=1$ only rockets at $\approx d_\text{LoS}$ due to the increase in the error probability. Therefore, an effectively coordinated multi-RAT WMN maximizes reliability and minimizes delay. 
 
%%%%%%%%%%%%%%%%%%%%%%%%%%%%%%%%%%%%%%%%%%%%
\section{Distributed consensus protocol}
\label{sec:protocol}
In this section, we describe our distributed consensus protocol. It is performed in the setup phase of the devices to form a WMN with a hierarchical architecture and is based on a set of rules that dictate the actions that must be performed to ensure convergence. Naturally, these rules are based, in turn, on the set of roles $\mathcal{H}=\{ \text{on}, \widehat{\text{CM}}, \text{CM}, \widehat{\text{CH}}, \text{CH}, \text{M}\}$ defined in the network , where $\widehat{\cdot}$ indicates a device that is not associated with a Master. These roles are updated in two phases. The first one is a neighbor discovery handshake and the second one is the distributed consensus. 

The setup phase of a device begins immediately after powering on, which may occur at any point in time. The hierarchy of the device is simply 'on' for a given period (e.g. calculated randomly) and then becomes a \wh{CH} if it was not discovered by a CH or Master. Analogously, a \wh{CH} becomes a Master after a certain period if it was not discovered by a Master. This ensures that at least one Master is elected at the end of the setup phase of all the devices.

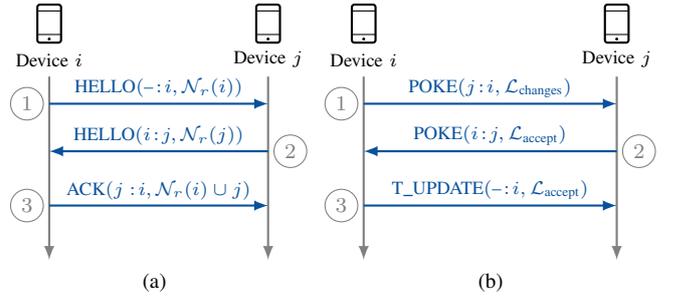
\begin{figure}
    \centering
    \subfloat[]{
   \begin{tikzpicture}[scale=0.9, block/.style={rectangle, minimum size=7mm,draw=gray, rounded corners=2pt},
   state/.append style={thin, gray, minimum size=12pt, inner sep=2pt}]
\tikzset{every node/.append style={font=\scriptsize}}
\def\x{2}
\def\y{5.2}
\pic[scale=1.2] at (\x,-0.6) {phone};
\pic[scale=1.2] at (\y,-0.6) {phone};

\draw [thick,->,gray](\y,-1.1)node[anchor=south, black]{Device $j$}--(\y,-3.8);
\draw [thick, ->,gray](\x,-1.1)node[anchor=south, black]{Device $i$}--(\x,-3.8);

\draw[->, Blues-K, thick] (\x,-1.5)node[state, left, xshift=-2pt]{$1$} -- (\y,-1.5)node[Blues-K, midway,above, inner sep=2pt]{HELLO$\left(\text{--}\! :\!i, \mathcal{N}_r(i)\right)$};
\draw[->, Blues-K, thick] (\y,-2.2)node[state, right, xshift=2pt]{$2$} -- (\x,-2.2)node[Blues-K, midway,above, inner sep=2pt]{HELLO$\left(i\!:\!j,\mathcal{N}_r(j)\right)$};

\draw[->, Blues-K, thick] (\x,-3)node[state, left, xshift=-2pt]{$3$} -- (\y,-3)node[Blues-K, midway,above, inner sep=2pt]{ACK$\left(j:\!i, \mathcal{N}_r(i)\cup j\right)$};
%\draw[->,Blues-K, thick] (\y,-3.8)node[state, right, xshift=2pt]{$4$}--(\x,-3.8)node[Blues-K,midway,above, inner sep=2pt]{HELLO$\left(i\! :\!\text{sym}\right)$};
\end{tikzpicture}\label{fig:ndisc_a}}\hfil
\subfloat[]{
\begin{tikzpicture}[scale=0.9, block/.style={rectangle, minimum size=7mm,draw=gray, rounded corners=2pt}]
\tikzset{every node/.append style={font=\scriptsize},
state/.append style={thin, gray, minimum size=12pt, inner sep=2pt}}
\def\x{2}
\def\y{5.7}
\pic[scale=1.2] at (\x,-0.6) {phone};
\pic[scale=1.2] at (\y,-0.6) {phone};

\draw [thick,->,gray](\y,-1.1)node[anchor=south, black]{Device $j$}--(\y,-3.8);
\draw [thick, ->,gray](\x,-1.1)node[anchor=south, black]{Device $i$}--(\x,-3.8);

%\phantom{\draw [->](\x,-1.1)--(\x,-4.6);}

\draw[->, Blues-K, thick] (\x,-1.5)node[state, left, xshift=-2pt]{$1$} -- (\y,-1.5)node[Blues-K, midway,above, inner sep=2pt]{POKE$\left(j\! :\!i,\mathcal{L}_\text{changes}\right)$};
\draw[->, Blues-K, thick] (\y,-2.2)node[state, right, xshift=2pt]{$2$} -- (\x,-2.2)node[Blues-K, midway,above, inner sep=2pt]{POKE$\left(i\! :\!j,\mathcal{L}_\text{accept}\right)$};
\draw[->, Blues-K, thick] (\x,-3)node[state, left, xshift=-2pt]{$3$} -- (\y,-3)node[Blues-K, midway,above, inner sep=2pt]{T\_UPDATE$\left(\text{--}\! :\!i,\mathcal{L}_\text{accept}\right)$};
\end{tikzpicture}
\label{fig:ndisc_b}}
\caption{(a) Neighbor discovery and (b) POKE message exchange to update the roles of the devices.}
\label{fig:ndisc}
\end{figure}

Immediately after becoming a \wh{CH} or Master, each device initiates active neighbor discovery through the short-range RATs or the long-range RAT, respectively. This is performed through any neighbor discovery algorithm that includes the three-message handshake illustrated in Fig.~\ref{fig:ndisc_a}. Specifically, a device $i$ initiates the active discovery by broadcasting a HELLO message through a RAT $r\in\mathcal{R}$, which includes a timestamp, its ID, and its open neighborhood $\mathcal{N}_r(i)$. 
Then, it waits for a predefined time window to receive the responses from nearby devices through the same RAT $r$. Any nearby device $j\notin \mathcal{N}_r(i)$ contend to access the wireless medium according to a random access (RA) protocol. A response to a HELLO message contains a timestamp, the ID $j$ of the device, and $\mathcal{N}_r(j)$. Then, $i$ confirms the reception of a successful response from a device $j$ by sending any type of acknowledgment (i.e., explicit or implicit) that includes $\mathcal{N}_r(i)\cup j$. 

A device $i$ initiates the distributed consensus phase through RAT $r$ immediately after concluding the neighbor discovery with the exact same RAT. Here, $i$ sends a POKE message as illustrated in Fig.~\ref{fig:ndisc_b} through RAT $r$ to a device $j\in\mathcal{N}_r(i)$. Each POKE message contains a timestamp, the ID of the transmitting device $i$, the ID of the target device $j$, and a list of suggested changes $\mathcal{L}_\text{changes}$. If the suggested changes comply with a predefined set of rules, device $j$ accepts the changes and sends a response to $i$. Then, $i$ transmits a topology update message T\_UPDATE to all devices whose state is affected by the changes. This message includes a timestamp, the ID of $i$, and the list of accepted changes. Then, device $i$ transmits POKE messages to the next $j\in\mathcal{N}_r(i)$.

The status of a device is defined by its role, but also by its CH and Master. Hence, we define the functions $m(\cdot)$ and $c(\cdot)$ which return the Master and the CH of a device. The set of rules defined by our protocol are used to create bilateral agreements between devices $i$ and $j$ to consistently change their status. The sets of rules are defined in Table~\ref{tab:short-rangerules} and Table~\ref{tab:long-rangerules} for the short-range RATs and the long-range RATs, respectively. It should be noted that not only the initial status of the devices, but also the device that initiates the handshake $i$ determines the resulting changes.

In the following Section, we provide simulation results that showcase the distinctive characteristics of our distributed consensus protocol for multi-RAT WMNs.

\begin{table}[t]
    \centering
    \scriptsize
    \astretch{1.2}
    \caption{Rules for the status changes during the exchange of POKE messages through the short-range RAT $r=1$.}
    \begin{tabular}{@{}llcl@{}}
    \toprule
         \multicolumn{2}{@{}c}{Initial roles} &  \multicolumn{1}{c}{Conditions} &  List of changes\\\cmidrule{1-2}
        \multicolumn{1}{@{}c}{$i$} & \multicolumn{1}{c}{$j$} \\\midrule
        \wh{CM} & on &  ---&  $j\leftarrow$\wh{CM}, $c(j)\leftarrow m$\\
        
        CM & on & --- & $j\leftarrow$CM, $c(j)\leftarrow m$, $m(j)\leftarrow n$\\ 
        
        CM & Master &  --- & $c(i)\leftarrow j$, $m(i)\leftarrow j$\\ 
        
        \wh{CH} & on &  --- & $j\leftarrow$\wh{CH}, $c(j)\leftarrow i$\\ 
        
         \wh{CH} & \wh{CH} &  --- & $j\leftarrow$\wh{CH}, $c(j)\leftarrow i$\\
         
         \wh{CH} & CH &  --- & $i\leftarrow$CM, $c(i)\leftarrow j$, $m(i)\leftarrow m$\\ 
        
        \wh{CH} & Master &  --- & $i\leftarrow$CM, $c(i)\leftarrow j$, $m(i)\leftarrow j$\\ 
         
         CH & on &  --- & $j\leftarrow$CM, $c(j)\leftarrow i$, $m(j)\leftarrow m$\\ 
        
        CH & CH &  $m(i)==m(j)$ & $j\leftarrow$CM, $c(j)\leftarrow i$\\ 
        
         CH & Master &  --- & $i\leftarrow$CM, $c(i)\leftarrow j$, $m(i)\leftarrow j$\\
         
         Master & any &  --- & $j\leftarrow$CM, $c(j)\leftarrow i$, $m(j)\leftarrow i$\\
      \bottomrule
      \vspace{0.5em}
    \end{tabular}
    \label{tab:short-rangerules}
\end{table}

\begin{table}[t]
    \centering
    \scriptsize
    \astretch{1.2}
    \caption{Rules for the status changes during the exchange of POKE messages through the long-range RAT $r=2$.}
    \begin{tabularx}{\columnwidth}{@{}Xlll@{}}
    \toprule
         \multicolumn{2}{@{}c}{Initial roles} &  \multicolumn{1}{c}{Conditions} &  List of changes\\\cmidrule{1-2}
        \multicolumn{1}{@{}c}{$i$} & \multicolumn{1}{c}{$j$} \\\midrule
        \wh{CH} & Master\hspace{-0.3em} &  $\mathcal{N}_2[i]\subseteq \mathcal{N}_2[j]$ &  $i\leftarrow$CH, $c(i)\leftarrow j$\\
        
        CH & Master\hspace{-0.3em} & $\mathcal{N}_2[i]\supset \mathcal{N}_2[j],m(i)==j$\hspace{-0.7em} & $i\leftarrow$Master, $j\leftarrow$CH, $m(j)\leftarrow i$\\ 
        
        Master & \wh{CH} &  $\mathcal{N}_2[i]\supseteq \mathcal{N}_2[j]$ & $j\leftarrow$CH, $m(j)\leftarrow i$\\ 
        
        Master & CH &  $\mathcal{N}_2[i]\subset \mathcal{N}_2[j], m(j)==i$\hspace{-0.7em} & $j\leftarrow$CH, $m(j)\leftarrow i$\\ 
        
         Master & Master\hspace{-0.3em} &  $\mathcal{N}_2[i]\supseteq \mathcal{N}_2[j]$ & $j\leftarrow$CH, $m(j)\leftarrow i$\\
         
         Master & Master\hspace{-0.3em} &  $\mathcal{N}_2[i]\subset \mathcal{N}_2[j]$ & $i\leftarrow$CH, $m(i)\leftarrow j$\\
      \bottomrule
    \end{tabularx}
    \label{tab:long-rangerules}
\end{table}

\iffalse      
{$i$} &  & \multicolumn{2}{c}{Device $j$} & &
         \multicolumn{2}{c}{Device $i$} &  &
         \multicolumn{2}{c}{Device $j$}\\\cmidrule{1-2}
         \cmidrule{4-5} \cmidrule{7-8} \cmidrule{10-11}  
         Role  & $m(i)$ & & Role & $m(j)$ & & Role & $m(i)$ & & Role & $m(j)$\\ 
         \midrule
           \\

         M & $i$ & & \wh{CH} & --- &
         $\mathcal{N}_1[i]\supseteq \mathcal{N}_1[j]$ &
         M & $i$ & & \cellcolor{Greys-D}CH  &\cellcolor{Greys-D}$i$ \\
         
         M & $i$ & & CH & $i$ &
         $\mathcal{N}_1[i]\subset \mathcal{N}_1[j]$ &
         \cellcolor{Greys-D}CH &\cellcolor{Greys-D}$j$ & & \cellcolor{Greys-D}M  &\cellcolor{Greys-D}$j$ \\
         
         M & $i$ & & M & $j$ &
         $\mathcal{N}_1[i]\supseteq \mathcal{N}_1[j]$ &
         M & $i$ & & \cellcolor{Greys-D}CH  &\cellcolor{Greys-D}$i$ \\
         
          M & $i$ & & M & $j$ &
         $\mathcal{N}_1[i]\subset \mathcal{N}_1[j]$ &
         \cellcolor{Greys-D}CH  &\cellcolor{Greys-D}$j$ & & M & $j$\\
\fi

%In the simple example of Fig.~\ref{fig:minimum_example} with MCo, finding the optimal broadcast strategy is not straightforward, even if the weight of the edges are known. For instance, assume $1$ is the source. To minimize the transmission latency, it is clear that edge $(1,2,2)$ must be used with some combination of long-range transmissions. 
%In the following, we describe three routing approaches to solve this problem.

%%%%%%%%%%%%%%%%%%%%%%%%%%%%%%%%%%%%%%%%%%%%%%%%%%%%%%%%%%%%%%%%%%%%%%%%%%%%%%%%
\section{Results}
\label{sec:results}

For the evaluation of our distributed consensus protocol we assume the devices are within a circular area $A=\pi r_A^2$ with radius $r_A$ according to a Poisson Point Process (PPP) $\Phi$ with intensity $\Lambda(A)=\E\left[N\right]$ where $N=\Phi(A)$ is the number of points in $\Phi\cap A$.
\begin{equation}
    p_{N}(n) = \frac{\left(\Lambda(A) \right)^n}{n!}e^{-\Lambda(A)}
\end{equation}

A discrete-event simulator was coded in Python for the analysis of our distributed consensus protocol. Each simulation begins at time $t=0$ and concludes when all devices have powered on and our distributed consensus protocol has reached the stable state where no more changes will occur. The number of simulation runs is sufficiently large to ensure that the relative margin of error for the presented results is less than $0.5$~\% at a $95$~\% confidence interval.

A particular device $i$ powers on at time $T_\text{on}^{(i)}\sim \mathit{Exp}(\lambda_\text{on})$. Then, the timers to transition to \wh{CH} and Master are set to $T_\text{CH}^{(i)}\sim \mathit{Exp}(\lambda_\text{CH})$ and $T_\text{M}^{(i)}\sim \mathit{Exp}(\lambda_\text{M})$, respectively. Hence, $i$ transitions to \wh{CH} at time $T_\text{on}^{(i)}+T_\text{CH}^{(i)}$ and to Master at time $T_\text{on}^{(i)}+T_\text{CH}^{(i)}+T_\text{M}^{(i)}$ if it is not discovered by a CH or Master. The values selected for the analysis are $\Lambda(A)=50$, $r_A=\{500, 1000,2000\}$~m, and $2\lambda_\text{on}=\lambda_\text{CH}=\lambda_\text{M}=1/5$ transitions per second; these are far greater than the time needed to transmit the HELLO, POKE, and T\_UPDATE messages to ensure consistency.

We selected two performance indicators to evaluate the benefits of implementing our distributed consensus protocol in a multi-RAT WMN w.r.t. a traditional routing protocol implemented in a WMN with $r=2$ only. The performance indicators are: 1) the probability of having only one device with the highest hierarchy and 2) the number of handshakes needed to setup the network. The first performance indicator is relevant because having a unique Master ensures that the resources in the network are correctly allocated and that the correct medium access protocol takes place. On the other hand, minimizing the utilization of $r=2$ is of utmost importance as most of the available frequency bands for this RAT are subject to duty cycle.

\begin{figure}
    \centering
    \subfloat[]{\includegraphics{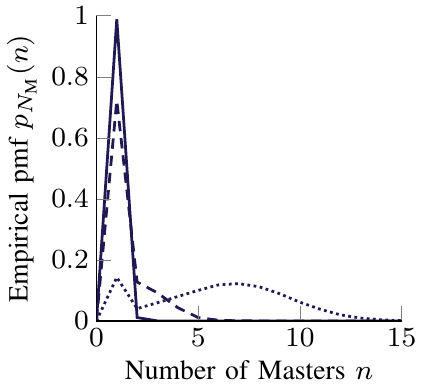}}
    \subfloat[]{\includegraphics{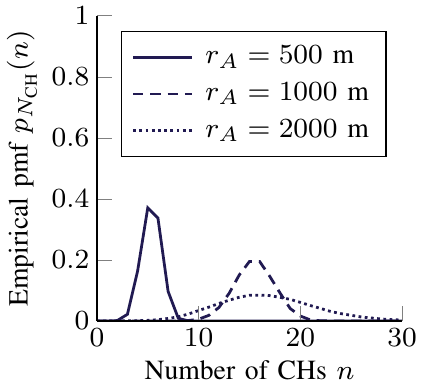}}\\
    \caption{Empirical pmf of the number of (a) Masters $p_{N_\text{M}}(n)$ and (b) CHs $p_{N_\text{CH}}(n)$ after setup.}
    \label{fig:num_roles}
\end{figure}

As a starting point, let $N_\text{M}$ and $N_\text{CH}$ be the RVs of the number of Masters and CHs in the network. Their empirical pmf is shown in Fig.~\ref{fig:num_roles}. It is important to observe from  Fig.~\ref{fig:num_roles} that a unique Master is selected whenever there exists a node $i$ that can communicate with the rest of the CHs -- and temporal Masters (i.e, those that transitioned due to a timer expiration but will change their roles after the distributed consensus) -- in the network. 

Let $V_2$ be the set vertices of $G_2(V_2,E_2)$, the random network graph with RAT $r=2$ at time $t$. Note that $V_2$ is the set of CHs and temporal Masters that were elected in a network with $N$ devices at an arbitrary time $t'<t$. A unique Master is elected from $V_2$ by our distributed consensus protocol at time $t$ if $\deg({G_2})=|V_2|-1$. For a specific vertex $i\in V_2$ we have
\begin{equation}
    \Pr\left[\deg(i)=|V_2|-1\right]=\prod_{\substack{j\in V_2,j\neq i}}p_{ij}^{(2)}\geq \prod_{\substack{j=1,j\neq i}}^{N}p_{ij}^{(2)}
\end{equation}
Therefore, the probability of having one Master is
\begin{equation}
    p_{N_\text{M}}(1)=%\Pr\left[\deg({G_2})=|V_2|-1\right]:=
    \Pr\left[\bigcup_{i\in V_2}\deg(i)=|V_2|-1\right],
\end{equation}
which decreases as $|V_2|$ increases. Consequently, reducing $|V_2|$ through the selection of CHs and CMs increases the probability of having a unique Master when compared to having a coordinator that communicates directly with each device in a WMN with $r=2$ only.

For the selected scenario, a unique Master is elected with probability $0.988$ for $r_A=500$~m, $0.723$ for $r_A=1000$~m, and $0.190$ for $r_A=2000$~m. 

Next, we illustrate the efficiency of our distributed consensus protocol w.r.t. neighbor discovery in a WMN with $r=2$ only. Let $X^{(\mathcal{R})}$ and $X^{(2)}$ be the minimum number of executions of a particular handshake that are needed to complete the network setup in a multi-RAT WMN and in a WMN with $r=2$ only. That is, to discover the neighborhood (HELLO) and to complete our distributed consensus protocol (POKE and T\_UPDATE). Note that direct communication between devices is needed to establish the links and to build trust. The minimum number of handshakes was selected as a performance indicator to focus on the efficiency of the protocol and not on the efficiency of its implementation. 

\begin{figure}
    \centering
    \includegraphics{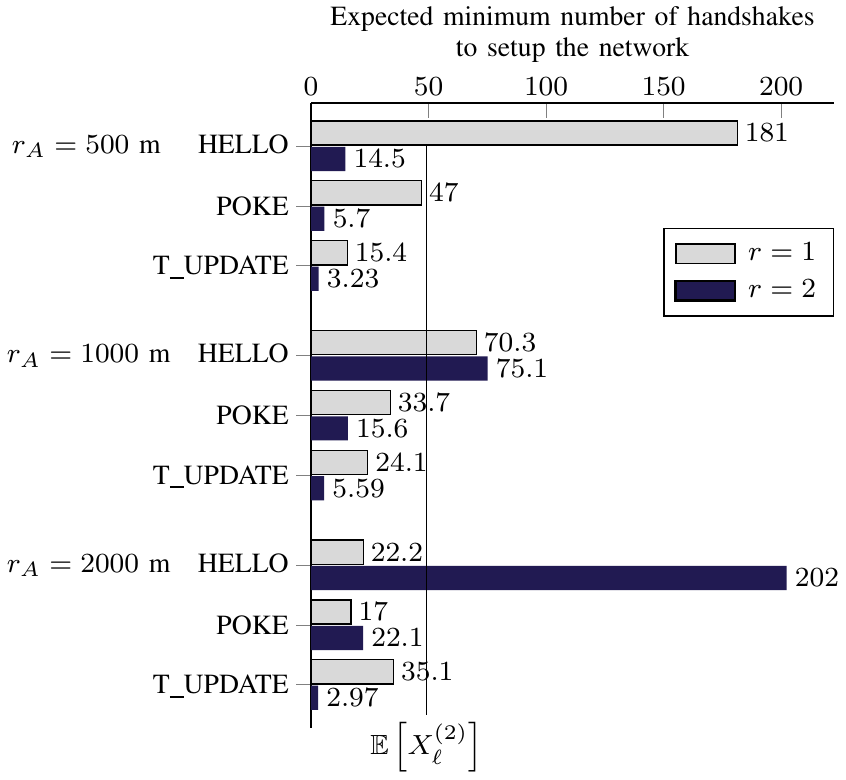}
    \caption{Expected minimum number of handshakes $\mathbb{E}\left[X^{(\mathcal{R})}\right]$ to reach the consensus with our distributed protocol.}
    \label{fig:num_messages}
\end{figure}

Fig.~\ref{fig:num_messages} shows $\mathbb{E}\left[X^{(\mathcal{R})}\right]$ for the HELLO, POKE, and T\_UPDATE in a multi-RAT WMN with our distributed consensus protocol.
As it can be seen, the utilization of $r=2$ is minimal for low values of $r_A$ and is mostly used for neighbor discovery. Naturally, the use of $r=2$ increases with $r_A$ due to the decrease in $|\mathcal{N}_1(i)|$ (number of short-range neighbors) for all $i$. 
The opposite effect is observed for $r=1$, for which our protocol does not affect the number of required HELLO handshakes. The decrease in the number of exchanges with $r=1$ is because $|\mathcal{N}_1(i)|$ decreases as $r_A$ increases.

Note that HELLO and POKE handshakes require a similar amount of resources. On the other hand, the resources needed to disseminate T\_UPDATE messages are higher due to the necessity to reach all the affected devices. Nevertheless, the number of T\_UPDATE messages that must be disseminated through $r=2$ is negligible when compared to the required number of neighbor discovery handshakes. 

The utilization of $r=2$ in a single-RAT WMN would be excessive if one handshake must be performed per link. We demonstrate this in the following by obtaining upper and lower bounds. 

A complete graph with $N$ vertices has $N(N+1)/2$ edges, whereas the minimum number of edges in a connected graph  with $N$ vertices is $N-1$. Therefore, for $N$ devices to discover each other directly in a complete graph, which occurs with high probability. for low values of $r_A$, $N(N+1)/2$ handshakes are needed to discover the neighborhood. On the other hand, a minimum of $N-1$ handshakes are needed to discover the neighborhood in a connected network. Building on this, and given $N$ has a Poisson distribution with intensity $\Lambda(A)=50$, for the neighbor discovery in a WMN with $r=2$ only we calculate the upper bound $\mathbb{E}\left[X_u^{(2)}\right]=1300$
%\begin{equation}
%   \mathbb{E}\left[X_u^{(2)}\right]= %\sum_{n=1}^{\infty}\frac{n(n+1)}{2} %p_N(n)=1300,
%\end{equation}
and the lower bound $\mathbb{E}\left[X_\ell^{(2)}\right]=49$ (shown in Fig.~\ref{fig:num_messages}).
%\begin{equation}
%   \mathbb{E}\left[X_\ell^{(2)}\right]= %\sum_{n=1}^{\infty}(n-1) p_N(n)=49.
%\end{equation}
Note these are in the order of $100\times$ and $3\times$ the number shown in Fig.~\ref{fig:num_messages} for $r_A=500$~m, respectively.

\balance
\section{Conclusions}
\label{sec:conc}

In this paper, we illustrated the benefits of multi-RAT WMNs and presented a distributed consensus protocol to achieve a hierarchical architecture and effectively coordinate the network. 

Our results emphasize that an effectively coordinated multi-RAT WMN increases the reliability and minimizes the latency of communication when compared to single-RAT WMNs. Please observe that coordination is not a specific overhead of multi-RAT WMNs. Instead, as observed by {\"{O}}zg{\"{u}}r et al.~\cite{Ozgur2007}, coordination is essential to maximize the performance of any interference-limited WMNs, regardless of the number of available RATs. 

On the other hand, our analyses show that our distributed consensus protocol ensures that all the devices have a coordinator (Master or CH) for all the available RATs and that the number of Masters is minimized. Having a unique Master in the network is relevant because it ensures that the communication is efficiently coordinated, that the resources are correctly allocated, and that the correct medium access protocol takes place. In addition, the unique Master can then implement advanced clustering techniques to maximize the performance of the network. Otherwise, the Masters must first coordinate themselves, which, in turn, complicates the coordination of the network.

A major concern in duty cycled frequency bands, such as the selected long-range RAT in the $868$~MHz, is air time efficiency. Results presented in Section~\ref{sec:results} show that our distributed consensus protocol is greatly efficient in this regard, even when compared to the lower band for neighbor discovery through this long-range RAT. 

 Other benefits of multi-RAT WMNs such as spatial and frequency reuse were not investigated due to the lack of space and are left for future work.

\bibliographystyle{IEEEtran}
\bibliography{Bibliography.bib}

\end{document}